\documentclass[aps,prb,preprint,amsfonts,showpacs,letterpaper]{revtex4}
\pdfoutput=1
\usepackage{graphicx}

\begin{document}

%\begin{CJK*}{UTF8}{bsmi}  % Use default fonts from CJK (see below)
\title{Obtaining correct orbital ground states in $f$ electron
systems using a nonspherical self-interaction corrected LDA+$U$ method}
\author{Fei Zhou}%(周非)}
\author{V. Ozoli\c{n}\v{s}}
\affiliation{Department of Materials Science and Engineering,
  University of California, Los Angeles, CA 90095} 

\date{\today}

\begin{abstract}
The electronic structure of lanthanide and actinide compounds is often
characterized by orbital ordering of localized
$f$-electrons. Density-functional theory (DFT) studies of such systems
using the currently available LDA+$U$ method are plagued by
significant orbital-dependent self-interaction, leading to erroneous
orbital ground states.  An alternative scheme that modifies the
exchange, not Hartree, energy is proposed as a remedy. We show
that our LDA+$U$ approach reproduces the expected degeneracy of $f^1$
and certain $f^2$ states in free ions and the correct ground states in solid
PrO$_2$. We expect our method to be useful in studying electronic excitations and entropies in $f$- and heavy-$d$ elements.
\end{abstract}

\maketitle
%\end{CJK*}

\section{Introduction}
Interesting physical phenomena associated with the strongly correlated
$f$-electrons in lanthanide and actinide compounds continue to attract lively
interest \cite{Pepper1991CR719,Dolg1996}.
%since the localized nature of the $f$ electrons causes most approximate 
%functionals to fail qualitatively due to inadequate treatment of
%strong on-site interactions between the $f$-electrons. 
%  REPLACED by FZ
Strong on-site interactions between the $f$-electrons in these materials 
present serious challenges to modern density-functional theory (DFT) 
based electronic-structure techniques, causing most approximate
functionals, such as the local density (LDA) or generalized gradient
approximation (GGA), to fail qualitatively.
To overcome the deficiencies of the LDA/GGA in studying $f$-element compounds, several recent studies
have employed the self-interaction-corrected LDA 
\cite{Perdew1981PRB5048}
e.g.\ in Refs.~\onlinecite{Svane2000IJQC799,Petit2003S498,Svane2007PRB115116},
the hybrid functional method
\cite{Becke1993JCP1372,Becke1993JCP5648}  in 
Refs.~\onlinecite{Kudin2002PRL266402, Prodan2005JCP14703,
Hay2006JCP34712,Prodan2006PRB45104,DaSilva2007PRB45121},
 or the dynamical mean-field theory (DMFT)
\cite{Metzner1989PRL324}
 in Refs.~\onlinecite{Pourovskii2007PRB235101,Jacob2008EL57009}.
The LDA+$U$ method  \cite{Anisimov1991PRB943} has emerged as a
well-established model to deal with strong electron correlations in 
$d$- and $f$-systems, combining high efficiency
with an explicit treatment of correlation within a Hubbard-like model
for the localized electrons. This method has been very
successful in transition metal oxides (for a review see
Ref.~\onlinecite{Anisimov2000}) and has yielded 
promising results for band gaps in $f$ systems
\cite{Larson2007PRB45114,DaSilva2007PRB45121,Tran2008PRB85123}.
However, systematic studies of its effectiveness
remain inconclusive, with issues of orbital ordering \cite{Hotta2006RPP2061}
and multiple self-consistent solutions attracting heightened attention
\cite{Shick2001JES753,Larson2007PRB45114,Jomard2008PRB75125,Amadon2008PRB155104, Ylvisaker2009PRB35103}.
%\cite{Larson2007PRB45114,Shick2001JES753-Jomard2008PRB75125-Amadon2008PRB155104}.
%%, pointing to greater 
%% importance of these effects in $f$ electron systems.

Here, we show that the currently popular versions of LDA+$U$, 
by Liechtenstein and co-workers \cite{Liechtenstein1995PRB5467} and by Dudarev and co-workers \cite{Dudarev1998PRB1505}, respectively, encounter serious
difficulties in $f$ systems due to large orbital-dependent 
self-interaction (SI) effects, which result in an unphysical
splitting of up to 0.4 eV between degenerate $f^1$ multiplets. Since
the SI errors (SIE) are typically larger than the crystal field (CF)
splitting energies, and comparable to the strength of the spin-orbit
coupling (SOC), they lead to qualitatively incorrect electronic ground
states in solids. We propose a new, orbital SI free form of the
LDA+$U$ method that leaves the LDA Hartree term intact and only
replaces the LDA exchange with the Hartree-Fock exchange. In our method,
the Hartree-Fock exchange term cancels the LDA self-interaction energy
to a high degree of accuracy, ensuring near-degeneracy of real- and
complex-valued orbitals in free ions and correctly reproducing the
$\Gamma_8$ ground state and $\Gamma_8 \rightarrow \Gamma_7$
excitation energies in the PrO$_2$ solid.
The accuracy of this functional is sufficient for evaluating
high-temperature electronic entropies of $f$ electron systems.

\section{Method and computational details}
All DFT calculations were carried out using the VASP package 
\cite{Kresse1996PRB11169,Kresse1999PRB1758} with projected augmented wave
(PAW) potentials \cite{Blochl1994PRB17953}, energy cutoff of 450 eV, and without any constraint symmetry or ionic relaxation.
For free ions, a 12 \AA cubic cell containing one ion and uniform compensating background charge were used. For the PrO$_{2}$ solid, we consider a primitive cell of the fcc supercell (lattice constant of 5.386
\AA \cite{Gardiner2004PRB24415}).
The term ``LDA+$U$'' is used irrespective of the $xc$ functional since
the LDA and GGA results are found similar. Each calculation was
initialized in a specific atomic orbital and self-consistently converged to either states very close to the
initial orbital with the results reported, or distinctly different states with lower energy. SOC was excluded from the calculations  unless its inclusion is stated explicitly to make realistic comparison with experiment. Finally, we fix the $U$ parameter in the LDA+$U$ method to 6 eV and leave the discussions of this choice to the end.

\subsection{Aspherical self-interaction error of LDA+$U$ for $f$-electrons}
\begin{figure}[htbp]
\includegraphics[width=0.8 \linewidth]{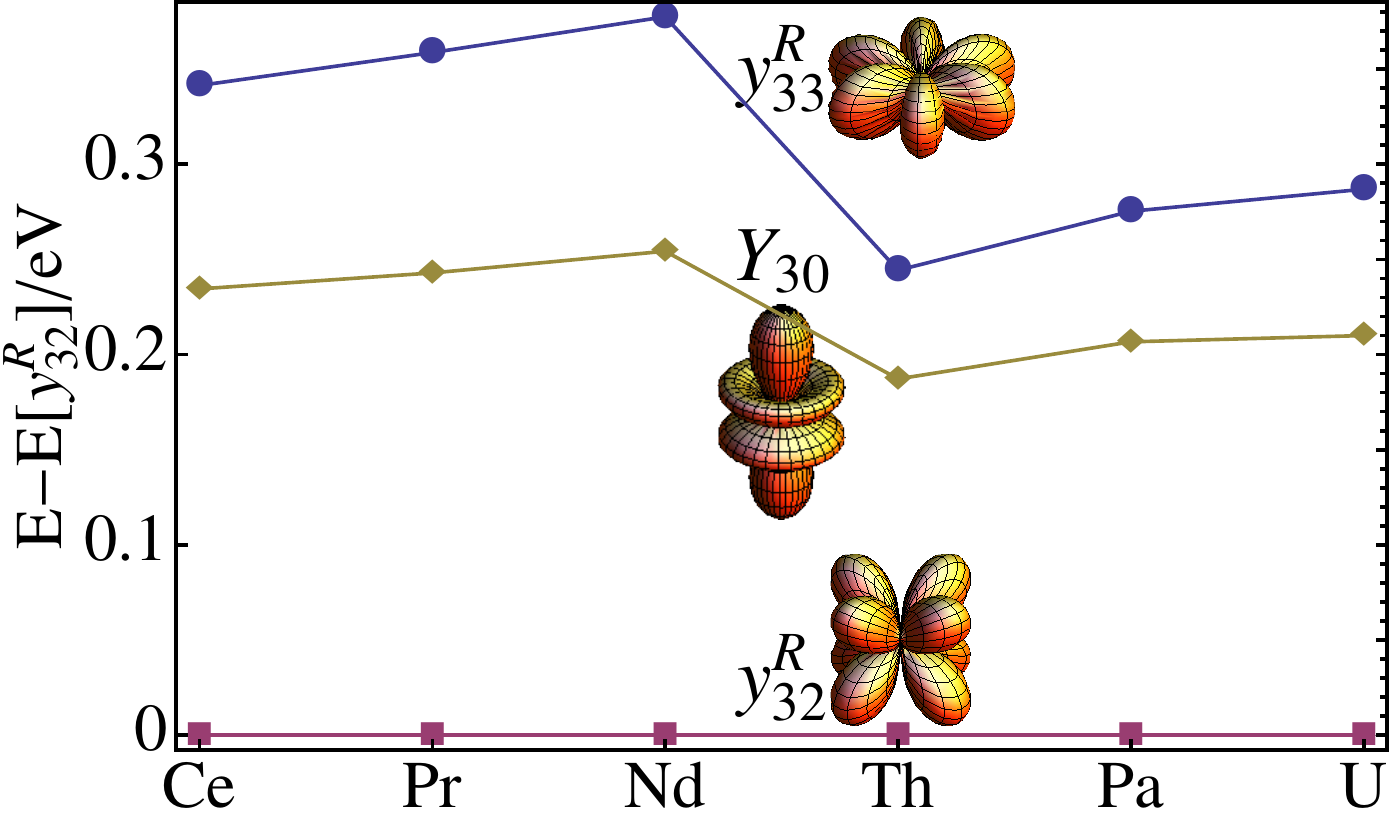}
\caption{LDA+$U$ total energies for different orbital filling of
 $f^1$ ions with the Liechtenstein scheme
  \cite{Liechtenstein1995PRB5467} relative to $y_{32}^{R}$.
\label{fig:E-ion}}
\end{figure}

We begin by showing that the conventional LDA+$U$ approach
fails to reproduce the degeneracy of different $| m \rangle$ orbitals of $f^{1}$ ions.
First consider real orbitals with angular dependence %given by
of real $y^R_{3m}=\sqrt{2} \Re Y_{3m}$ without spin-orbit effects to simplify
the presentation of our method. Complex orbitals and SOC are discussed later. 
Fig.\ \ref{fig:E-ion} shows the energies of
different $y^{R}_{3m}$ orbitals (with the exception of
$y^R_{31}$, which converges to $y^R_{32}$) in several lanthanide and actinide ions
calculated using the LDA+$U$ scheme of Liechtenstein {\it et al.\/} \cite{Liechtenstein1995PRB5467}
with $J=0.5$ and $U=6$~eV.
Contrary to the expected degeneracy, the energies of the
different $y^R_{3m}$ orbitals differ substantially, up to 0.4 eV, and the
$y^R_{31}$ orbital was found unstable and converged to $y^R_{32}$. 
Varying $J$ between 0 (i.e.\ the Dudarev scheme \cite{Dudarev1998PRB1505}) $\sim 1$ eV 
changes the results by only a few meV.

The above results demonstrate that the conventional LDA+$U$ approach commits large errors of up to 0.4 eV/electron in the  predicted relative orbital energies of $f$ electrons. To understand the reasons for the unphysical splitting of the $f^1$ states, we examine the conventional LDA+$U$ total energy functional\cite{Anisimov1991PRB943}:
\begin{eqnarray} 
E ^{\mathrm{LDA}+U} = E ^{\mathrm{LDA}} + E_U - E_{\mathrm{dc}},
\label{eq:lda+u} 
\end{eqnarray}
where the LDA description of the on-site interaction,
approximately represented by the so-called double-counting term
$E_{\mathrm{dc}}$, is replaced with a Hubbard-like $E_U$. The
latter is essentially the Hartree-Fock energy,
expressed in a rotationally invariant form by Liechtenstein {\it et  al.\/} 
\cite{Liechtenstein1995PRB5467} as a sum of the Hartree (H) and
exchange (X) terms, $E_U = E_{\mathrm{H}} + E_{\mathrm{X}}$, where
\begin{eqnarray} 
E_{\mathrm{H}}  &=& \frac12 \sum_{ \{m\} }  
 \langle m, m'' | V_{\mathrm{ee}} | m' ,m''' \rangle n_{m m'} n_{m'' m'''}, 
\label{eq:EHartree} \\
 E_{\mathrm{X}} &=& - \frac12 \sum_{ \{m\} , \sigma}  
\langle m, m'' |  V_{\mathrm{ee}} | m''' , m' \rangle  
n_{m m'}^\sigma n_{m'' m'''}^{\sigma}.  
\label{eq:EFock} 
\end{eqnarray}
The on-site density matrix $n^\sigma_{m m'}$
is obtained by projecting the Kohn-Sham orbitals $\psi_{\alpha}^{\sigma}$ of occupancy $f_{\alpha}^{\sigma}$ onto atomic states $| nlm(m') \rangle$
\begin{eqnarray}
n^\sigma_{m m'} &=& \sum_{\alpha} f_\alpha^\sigma \langle \psi_\alpha^\sigma | nlm' \rangle 
 \langle nlm | \psi^\sigma_\alpha \rangle,
\end{eqnarray}
while the Slater integrals 
$\langle m m'| V_{\mathrm{ee}}| m''m'''\rangle$ are evaluated in terms
of the Gaunt coefficients and 
the screened Coulomb $U$ and exchange $J$ parameters (the diagonal
$m=m'=m''=m'''$ terms are given in Table~\ref{tab:exchange-analytical-LSD}).
A simplified version by Dudarev {\it et al.} \cite{Dudarev1998PRB1505}
adopts the $J=0$ limit. States with real $n^\sigma_{m m'}$ are referred to as ``real''. 

\begin{table}[htbp] 
\begin{ruledtabular}
\begin{tabular}{|c|ccccccc|}
 	&  $y^R_{l3}$ & $y^R_{l2}$ 		& $y^R_{l1}$ 	&  $Y_{l0}$   & $Y_{l1}$ & $Y_{l2}$ & $Y_{l3}$ \\ \hline
$l$ 	& \multicolumn{7}{c|}{Value of $a$ in $E_{\mathrm{H}}=\langle mm |V_{ee}|mm \rangle/2=U/2+ a J$} \\ \hline
1 	&			& 		&0.4 		&0.4 &  0.1 & & \\ 
2 	&			& 0.571 	&0.571 	&0.571  & 0.186   &  0.358 & \\ 
3	& 	0.880  	& 0.422  	& 0.807  	& 0.716 & 0.332  & 0.194 &  0.696 \\ 
\hline
$l$ 	& \multicolumn{7}{c|}{Value of $a$ in LSD exchange $E_x^{\mathrm{LSD}} = -a K$} \\ \hline
1 	&			& 		&0.409 		& 0.409 &  0.364 & & \\ 
2 	&			& 0.364 & 0.364 	& 0.356  &  0.324  & 0.324 & \\ 
3	& 	0.339 & 0.328 & 0.335 & 0.323 & 0.298 & 0.292 & 0.302 \\ 
\end{tabular}
\end{ruledtabular}
\caption{
Hartree energy $E _{\mathrm{H}}$, Eq.~\ref{eq:EHartree},  and
LSD exchange energy $E_x^{\mathrm{LSD}}$, Eq.~(\ref{eq:exchange-LSD}),
for one $l$-electron in orbitals with real ($y^R_{lm}$,$Y_{l0}$) and complex
($Y_{lm}$ for $m>0$) angular wavefunctions.
\label{tab:exchange-analytical-LSD}} 
\end{table}

For a free $f^{1}$ ion, the Hartree-Fock
energy $E_U$ in Eq.~\ref{eq:lda+u} naturally vanishes,  while
$E_{\mathrm{dc}}$ in the Liechtenstein and Dudarev schemes 
depends only on the number of electrons, $N^\sigma=\sum_m n_{mm}^\sigma$, and
not on the type of the occupied orbital. Therefore, Eq.~\ref{eq:lda+u} becomes 
\begin{eqnarray*} 
E ^{\mathrm{LDA}+U} 
%= E _{\mathrm{H}} + E ^{\mathrm{LDA}}_{x} + \mathrm{const} 
%\approx E _{\mathrm{H}}  [\{ n \}] + \mathrm{const} .
=  E ^{\mathrm{LDA}}   + \mathrm{const} \approx E _{\mathrm{H}} +  \mathrm{const}.
\label{eq:E-one-electron}
\end{eqnarray*}
In the above approximation we assumed 1) the LDA exchange is not sensitive to orbital filling (more on this later) and 2) the Hartree energy difference comes mainly from the on-site Hartree term $E _{\mathrm{H}}$ of eq.~\ref{eq:EHartree}.
 The resulting error in the relative orbital energies is then
entirely due to the orbital-dependence of the SIE of the LDA, which is reflected in $E _{\mathrm{H}}$. To see the validity of our argument, we list in
Table~\ref{tab:exchange-analytical-LSD} the on-site $E _{\mathrm{H}}$
calculated from eq.~\ref{eq:EHartree} for atomic orbitals; 
these expressions are expected to closely approximate 
%the Hartree SI term of the LDA for localized orbitals.
% REPLACED by FZ
the SI for localized orbitals in the LDA+$U$.
Even though $E_{\mathrm{H}}$ is identical for all real
$p$ or $d$ orbitals, it is orbital-dependent for $f$ multiplets, and
in all cases splits the SI energies of real {\it vs.\/} complex orbitals. The 
predicted ordering of 
$E_{\mathrm{H}}$
is $y_{32}<
y_{30} < y_{31} < y_{33}$, in agreement with the LDA+$U$ results shown
in Fig.~\ref{fig:E-ion}, demonstrating
that the  unphysical splitting of $f^1$ states in conventional
LDA+$U$ is due to orbital-dependent SIE.
Note that with
real orbitals the problem of orbital-dependent SIE does not affect $p$ or $d$ electrons.
We will show later that complex $p$ and $d$ orbitals are affected. 
According to Table~\ref{tab:exchange-analytical-LSD}, the SIE is
proportional to $J$; for typical values of $J$ in the range of
$0.1$ to $1$ eV, it is comparable to or even larger than other important
on-site effects, such as CF and SOC, which can lead to qualitatively
incorrect predictions of electronic ground states in solids by the current LDA+$U$ methods.
These deficiencies of the conventional LDA+$U$ approach can be
traced back to its treatment of the Hartree and exchange energies.
The LDA+$U$ approach replaces the LDA Hartree energy
with an on-site model expression $E_{\mathrm{H}}$ given by
Eq.~\ref{eq:EHartree}. Even though the $E_{\mathrm{H}}$ term 
is capable of reproducing the correct orbital energetics, the 
LDA+$U$ double-counting energy $E_{\mathrm{dc}}$ is 
{\it orbital-indepedent\/} and fails to properly account for the
orbital-dependence of the LDA SIE in open-shell
systems. Similar considerations 
hold for the orbital-dependence of the LDA exchange energy, which
is mainly sensitive to the choice of real {\it vs.\/} complex orbitals
(see Table ~\ref{tab:exchange-analytical-LSD}); this factor acquires
importance in systems with strong SOC, when the orbitals with a
definite value of the total angular momentum $J$ are necessarily
complex.
% For the exact DFT
% functional, the SI energies in the Hartree and exchange terms would cancel,
% leading to correct reproduction of orbital energetics.
% The underlying 
% physical reason for the spurious splitting in Fig.~\ref{fig:E-ion} is
% the orbital-dependence of the LDA SIE, which is not properly removed
% in the current LDA+$U$ approaches.

\subsection{Reformulated LDA+$U$}
To correct the orbital-dependent SIE in the Hartree and exchange
terms, we propose a new formulation of the LDA+$U$ method by modifying
only the exchange term of the LDA:
\begin{eqnarray} 
E ^{\mathrm{LDA}+U} = E ^{\mathrm{LDA}} + E_{\mathrm{X}} - E_{\mathrm{dcX}},
% =E ^{\mathrm{LDA}} +\Delta E . 
\label{eq:newlda+u}
\end{eqnarray}
where the orbital-dependent Hartree-Fock exchange
$E_{\mathrm{X}}$ of Eq.~(\ref{eq:EFock}) contains a term that
approximately cancels the SIE in the LDA Hartree energy; the
remainder of the LDA Hartree energy is exact by definition and
therefore left unmodified in our approach.
The exchange double-counting term $E_{\mathrm{dcX}}$ accounts for the
LDA exchange energy and is given by a linear
combination of the exchange double-counting in the Liechtenstein scheme
and the on-site local-spin-density (LSD) exchange:
\begin{eqnarray}
%E_{\mathrm{dcX}} &=& (1-c) E_{\mathrm{dcX}}^{\mathrm{FLL}} + c E_\mathrm{X}^{\mathrm{LSD}} \\
%E_{\mathrm{dcX}} ^{\mathrm{FLL}} &=& \frac12 \sum_{ \sigma} [  U N^\sigma + J N^{\sigma} (N^{\sigma} -1)] \\
E_{\mathrm{dcX}} &=& -\frac{1-c}{2} \sum_{ \sigma} [  U N^\sigma + J N^{\sigma} (N^{\sigma} -1)] + cE_\mathrm{X}^{\mathrm{LSD}}, \label{eq:dcX}\\ 
% VO Please check - I introduced parentheses around rho^sigma
E_\mathrm{X}^{\mathrm{LSD}} &=& - \frac{3}{2} \left(\frac{3}{4 \pi}\right)^{1/3}  \sum_\sigma \int d^3r (\rho^{\sigma})^{4/3} \nonumber \\
%&=& - \frac{3}{2} \left(\frac{3}{4 \pi}\right)^{1/3} \sum_\sigma \int d^3r \left[n_{mm'}^{\sigma} \bar{\psi}_m(r) \psi_{m'} \right] ^{4/3} \nonumber \\
&=& - \frac{3}{2} \left(\frac{3}{4 \pi}\right)^{1/3} \sum_\sigma \int R_{l}^{8/3}(r) r^2 dr d\Omega  \left[ n_{mm'}^{\sigma} \bar{Y}_{lm} (\Omega) Y_{lm'} (\Omega) \right] ^{4/3} \nonumber \\
&=& -  \left( \frac{ 4 \pi}{2l +1} \right ) ^{1/3} \frac{K}{2} \sum_\sigma \int d\Omega \left[ n_{mm'}^{\sigma}  \bar{Y}_{lm} Y_{lm'}  \right] ^{4/3} , \nonumber\\
&=& - \left( \frac{ 4 \pi}{2l +1} \right ) ^{1/3} \frac{K}{2} \sum_\sigma \int d\Omega  \left[\tilde{\rho}^\sigma(\Omega)\right] ^{4/3} 
\label{eq:exchange-LSD}
\end{eqnarray}
where $c$ is the interpolation coefficient,
%between the exact exchange in the fully localized limit (first term
%in Eq.~(\ref{eq:dcX}) and the LSD exchange E_\mathrm{X}^{\mathrm{LSD}},
$\rho^{\sigma}$ is the charge density of spin component $\sigma$,
which can be obtained from the on-site occupation matrix $n_{mm'}^{\sigma}$ as well as radial function $R_{l}(r)$ and spherical $Y_{lm}(\Omega)$,
$K$ is the
% VO is ``screened'' here correct?
LSD exchange strength parameter, and  $\tilde{\rho}$ represents the angular part of $\rho$. Only the $E_\mathrm{X}^{\mathrm{LSD}}$ term in
Eq.~(\ref{eq:dcX}) is orbital-dependent. The linear interpolation 
%in Eq.~(\ref{eq:dcX}) 
is conceptually similar to hybrid functional
approaches and serves the purpose of subtracting the
orbital-dependence of the LDA exchange energy.
The potential corresponding to the correction energy $E_{\mathrm{X}} - E_{\mathrm{dcX}}$, obtained by differentiating with respect to the on-site density matrix $n_{mm'}$, is then 
\begin{eqnarray}
\Delta V^\sigma_{m m'} &=&  \frac{2c}{3} \left( \frac{ 4 \pi}{2l +1} \right ) ^{1/3} K \int d\Omega \left[\tilde{\rho}^\sigma(\Omega)\right] ^{1/3} \bar{Y}_{lm}  Y_{lm'} \nonumber \\
&+&(1-c)  (\frac{U-J} {2} + n^\sigma J) \delta_{m m'} - \langle m, m'' | V_{ee} | m''' m' \rangle   n_{m'' m'''}^{\sigma} 
\end{eqnarray}
It is possible to reduce the number of independent parameters by requiring that $E_{\mathrm{X}}-E_{\mathrm{dcX}}$ vanishes for full
$l$-shells ($n^{\uparrow}_{mm'}=n^{\downarrow}_{mm'}=\delta_{mm'}$),
$$
E_{\mathrm{X}}-E_{\mathrm{dcX}}= -c(2l+1)(U+2lJ)+ c(2l+1)K =0,
$$
which gives 
\begin{eqnarray}
K=U + 2lJ.
\end{eqnarray}
The main advantage of
Eqs.~(\ref{eq:newlda+u})-(\ref{eq:exchange-LSD}) is that the
LDA self-interaction energy is canceled by the corresponding exchange
term in $E_{\mathrm{X}}$. As a result, the proposed method is
self-interaction free to high accuracy.

\section{Results and discussions}
\begin{figure}[htbp]
\includegraphics[width=0.8 \linewidth]{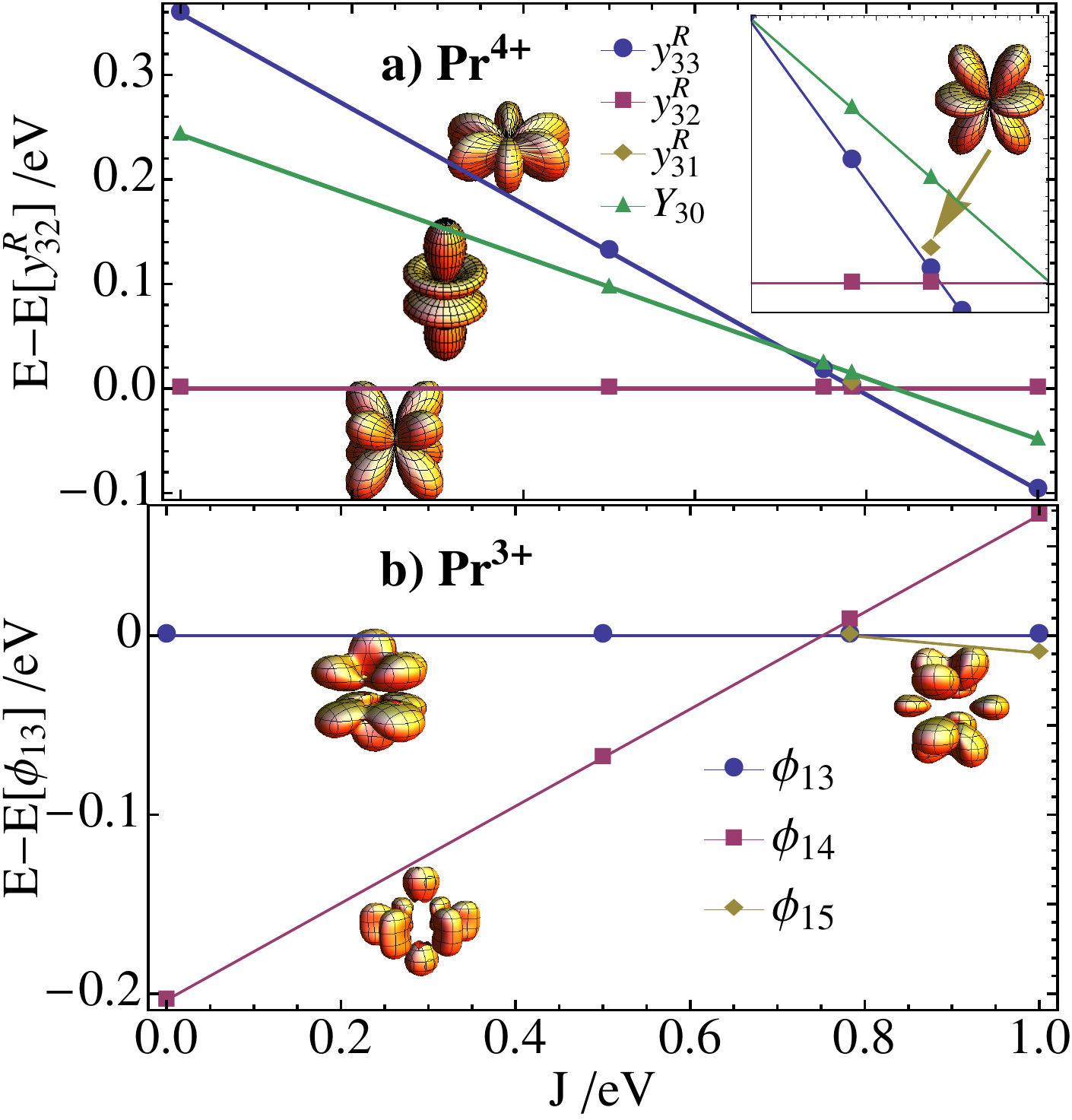}
\caption{LDA+$U$ energy of the Pr$^{3+}$ and Pr$^{4+}$ ions as a
  function of $J$ for $c=0$ calculated with our method. a) $f^1$ in
  orbitals $y^I_{3m}$, with the optimal $J$ region magnified in the
  inset;  b) $f^2$ in three degenerate (in terms of $E_U$)
  two-electron states.
} \label{fig:Pr-EvsJ-U6}
\end{figure}
\begin{figure}[htbp]
\includegraphics[width=0.8 \linewidth]{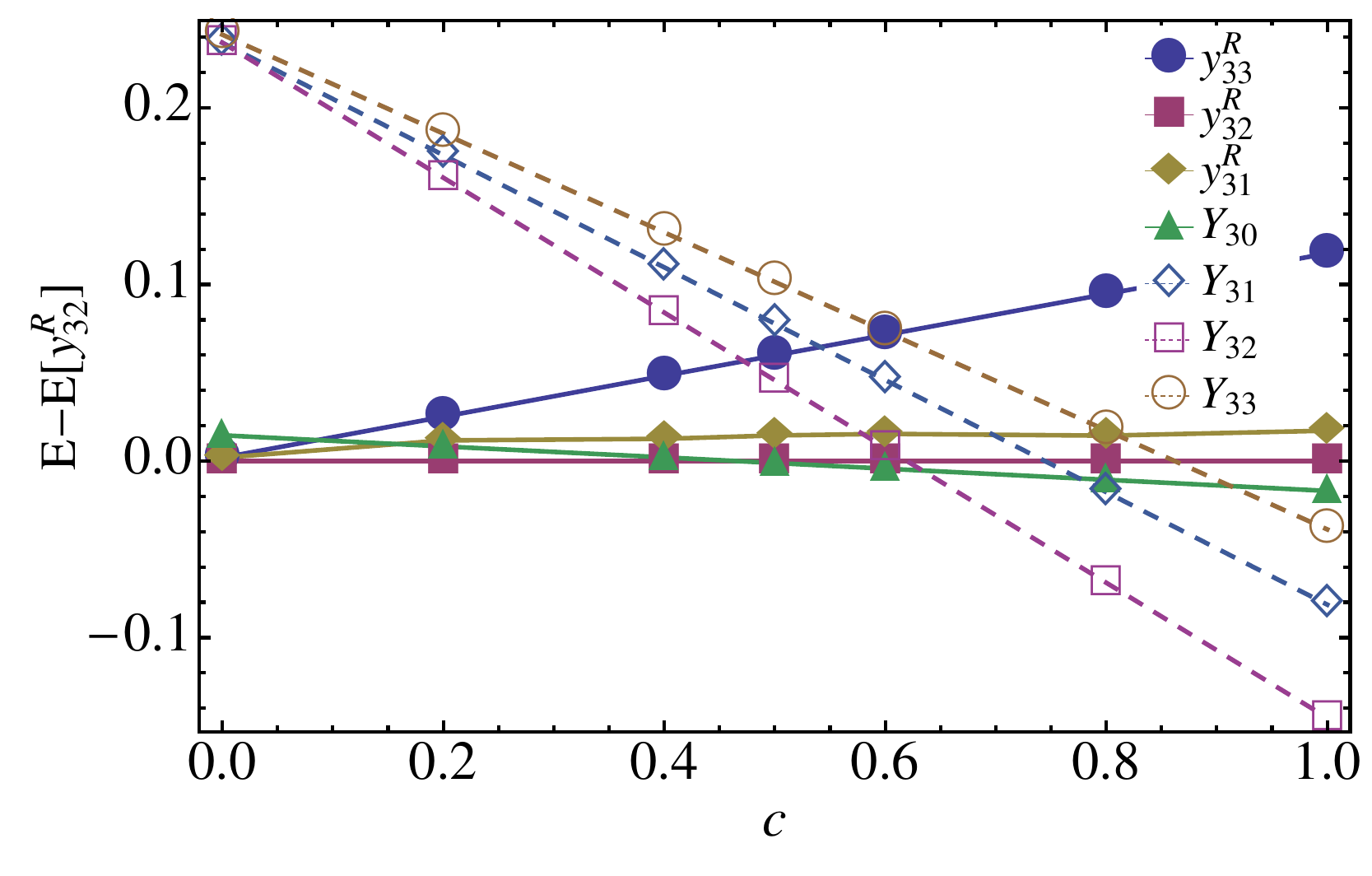}
\caption{LDA+$U$ energy of the Pr$^{4+}$ ion as a function of $c$ at fixed
  $J^{\mathrm{o}}=0.783$ eV calculated with our scheme, including
  both real- and complex-valued (no SOC) orbitals.
} \label{fig:Evsc}
\end{figure}
In this section, we analyze the parameter dependence of the proposed method and then presents results for the example of PrO$_{2}$ solid.

\subsection{Determination of parameters to remove aspherical SIE}
We demonstrate orbital degeneracy for free
Pr ions with one and two $f$-electrons. Figure~\ref{fig:Pr-EvsJ-U6}a
displays the energy of Pr$^{4+}$ in real atomic orbitals calculated
with our method (assuming $c=0$) as a function of the exchange parameter
$J$. At $J=0$, a splitting of more than 0.3 eV is found, similar to
the behavior of the original LDA+$U$ in Fig.~\ref{fig:E-ion}. The
splitting is reduced by increasing $J$ and at the optimal
value of $J^{\mathrm{o}}=0.783$~eV, it is less than 40~meV, 
i.e.,\ the four real orbitals $y^R_{3m}$ are almost degenerate. 
The $y^R_{31}$ orbital can only be stabilized in the
vicinity of $J^{\mathrm{o}}$, relaxing otherwise to the more stable $y^R_{32}$ or
$y^R_{33}$. Hence, just one point for $y^R_{31}$ is shown in the
inset of Fig.~\ref{fig:Pr-EvsJ-U6}a.
 
The energy of the Pr$^{3+}$ ion ($f^{2}$) is shown in
Fig.~\ref{fig:Pr-EvsJ-U6}b (also at $c=0$).
% the ground state is $^3H$
%by Hund's first two rules if spin-orbit coupling is ignored. The $S=1$
%ground state can be found by minimizing $E_U[n^\sigma_{m m'}]$
%with constraint $N^\uparrow=2$, $N^\downarrow=0$. 
Consider three distinct $f^2$ states with $S=1$ and degenerate Hartree-Fock energy
$E_U$.
Using the basis defined by real-valued spherical harmonics,
%$\{\dots, y^I_{lm}=\sqrt{2} \Im Y_{l-m}, \ldots, Y_{l0},
%\dots, y^R_{lm}=\sqrt{2} \Re Y_{lm}, \dots \}$
$\{ y^I_{l|m|}=\sqrt{2} \Im Y_{l|m|} (-l \le m < 0$), $Y_{l0}$,
%$y^R_{lm}=\sqrt{2} \Re Y_{lm}$ ($0 < m \le l$)
$y^R_{lm} (0 < m \le l)\}$
(shown for $l=3$ in Fig.~\ref{fig:energy-levels-WF}b), the first of
these states, designated by $\phi_{13}$, has electrons in orbitals
$y^I_{31}$ and $y^I_{33}$, or $n^\sigma_{mm'}=0$ except
$n^\uparrow_{11}=n^\uparrow_{33}=1$, while the other two $f^{2}$ states,
designated by $\phi_{14}$ and $\phi_{15}$,
% have one electron in
%in $y^I_{31}$ and the other in $Y_{30}$ and $y^R_{31}$ orbitals,
%respectively.
% ; the other two $f^{2}$ states
%are $\phi_{14}$ and $\phi_{15}$, 
correspond to
$n^\uparrow_{11}=n^\uparrow_{44}=1$ and
$n^\uparrow_{11}=n^\uparrow_{55}=1$, respectively. 
Their angular
wavefunctions are shown in Fig.~\ref{fig:Pr-EvsJ-U6}b. Similar to
the $f^{1}$ case, the energy splitting is large at
$J=0$ and gets reduced to less than 30 meV at the
optimal value $J^{\mathrm{o}}$. Note that $\phi_{15}$ can be
stabilized only for $J \gtrsim J^{\mathrm{o}}$.

So far, we have used $c=0$, assuming that the LSD
exchange functional is insensitive to the orbital and can be ignored. 
The lower part of Table~\ref{tab:exchange-analytical-LSD}  proves this
assumption for the real orbitals: $E_{\mathrm{X}}^{\mathrm{LSD}}$
varies by less than $0.02
K$. However, Table~\ref{tab:exchange-analytical-LSD} also shows that 
$E_{\mathrm{X}}^{\mathrm{LSD}}$ of complex orbitals is substantially
lower (by $\sim 0.3 K$), indicating a large lowering of the
exchange energy in states with nonzero orbital current.
Since $E_{\mathrm{X}}^{\mathrm{LSD}} \sim - \rho^{4/3}$ is concave, it favors inhomogeneous charge distributions (such as real orbitals compared to complex ones) and therefore the LDA exchange energies in Table~\ref{tab:exchange-analytical-LSD} of real orbitals are lower than those for complex orbitals.
 The difference  may play an important role in systems with strong
SOC, when the resulting electronic states are complex combinations of
real $y_{lm}$'s 
%with a definite value of the total angular momentum $J=L \pm S$.
% VO I am not sure I understand what is meant by this ...
with the orbital angular momentum unsuppressed.
In Fig.~\ref{fig:Evsc}, we show the dependence of the energies of real-
and complex-valued orbitals for Pr$^{4+}$ on the mixing coefficient $c$
%in front of the orbital-dependent double-counting
%$E_{\mathrm{X}}^{\mathrm{LSD}}$ 
in Eq.~(\ref{eq:exchange-LSD}), using the optimal
value of the exchange parameter, $J^{\mathrm{o}}$. It is seen that
at $c=0$, the energies of real and complex orbitals differ
by more than $0.2$~eV due to their different LSD exchange, and
the spurious splitting is 
%reduced by increasing the weight
%$c$ of the orbital-dependent double-counting and 
minimized to approximately 70 meV at the optimal $c\approx 0.6$.

In our approach, the $J$ and $c$ parameters are {\it a priori}
determined by the physical requirement of degeneracy once the $U$
parameter is given (6 eV in this work). They hardly change
when $U=4$ eV is used, suggesting that our method is relatively
insensitive to the choice of $U$.

\subsection{Eigenstates of PrO$_{2}$ without SOC}
\begin{table}[htbp] 
\begin{ruledtabular}
\begin{tabular}{|c|c|c|c|}
 			& $Y_{30} \ (t_{1u})$	&$y^R_{32} \ (t_{2u})$	&$y^I_{32}\ (a_{2u})$  \\ \hline
\multicolumn{4}{|c|}{Energy eigenvalue in cubic CF  (arbitr.\ unit)} \\ \hline
	& \bf{-3}	& 1	& 6 \\ \hline
\multicolumn{4}{|c|}{LDA+$U$ energy in different schemes (eV)} \\ \hline
Liechtenstein    & \bf{-23.848}         & \bf{-23.843}       &-23.488 \\ 
Dudarev         &-23.693          & \bf{-23.877}       &-23.458 \\
This work       & \bf{-24.260}            &-24.128       &-23.834 
\end{tabular}
\end{ruledtabular}
\caption{Comparison of the LDA+$U$ energy of PrO$_2$ and crystal field
  eigenvalues. Ground state energy is given in bold. \label{tab:PrO2}}
\end{table}
Finally, we demonstrate the advantages of our method for extended
solids by considering PrO$_2$ in the cubic fluorite structure.
The Pr$^{4+}$ ion is coordinated by eight oxygen atoms in a
cube. Figure~\ref{fig:energy-levels-WF} shows the $f^1$
energy level splitting scheme in the presence of cubic CF and
SOC. Without SOC, the cubic CF splits the $f^1$
states into the $t_{1u}$ ground state and $t_{2u}$, $a_{2u}$ excited
states (see Fig.~\ref{fig:energy-levels-WF}a,c). Table~\ref{tab:PrO2}
lists the CF eigenvalues of these states (small 6th-order CF
ignored), and the calculated LDA+$U$ energies using the conventional
approaches and our new scheme at the optimal values of $J=0.783$~eV
and $c=0.6$. The conventional schemes predict
orbital enegies that deviate dramatically from the expected CF
order: the Liechtenstein approach predicts almost degenerate 
$t_{1u}$ and $t_{2u}$, while $t_{2u}$ is the ground state in the Dudarev
method. In contrast, our new method
successfully finds the correct $t_{1u}$ ground state. 

\subsection{Eigenstates of PrO$_{2}$ with SOC}
\begin{figure}[htbp]
\includegraphics[width=0.85 \linewidth]{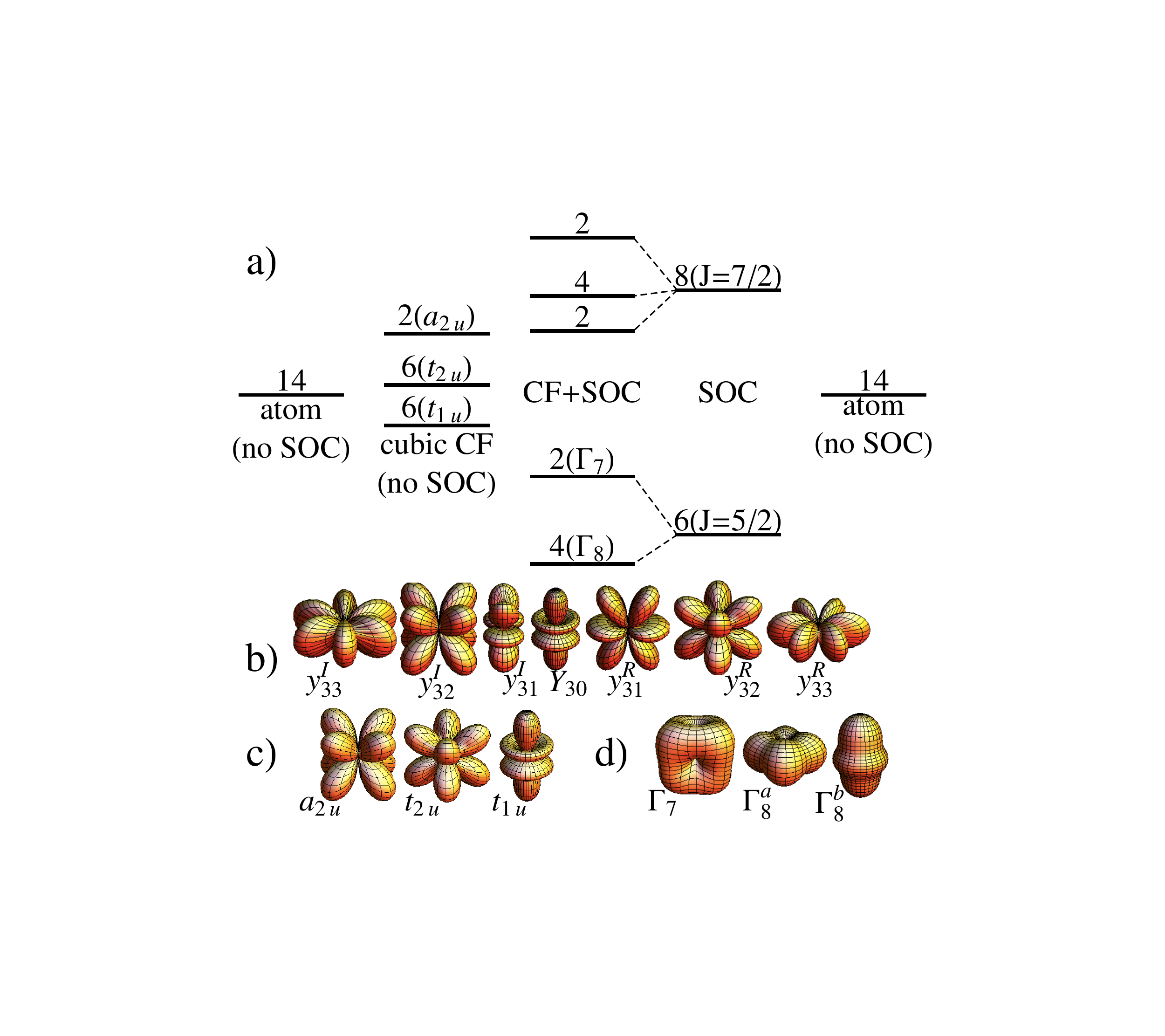}
\caption{a) Schematics of $f^1$ energy levels (with multiplicity)
  split by SOC and cubic CF. The angular wavefunctions are shown
  for b) real-valued atomic orbitals $\Psi_3$, CF eigenstates c)
  without and d) with SOC.}
\label{fig:energy-levels-WF}
\end{figure}
The physics of orbital ordering in $f$ systems is affected by
strong relativistic effects \cite{Hotta2006RPP2061}, 
necessitating the inclusion of SOC 
to make direct comparisons with experiment.
Including SOC, our method predicts
that the energies of the CF-degenerate $\Gamma_8^a$ and $\Gamma_8^b$, and
the excited $\Gamma_7$ states in PrO$_2$
(Fig.~\ref{fig:energy-levels-WF}d) are 0 (reference), 69 and 142 meV, 
respectively. The spurious 69 meV splitting between the two degenerate
$\Gamma_8$ states is consistent with the accuracy shown in
Fig.~\ref{fig:Evsc}. Neglecting Jan-Teller lattice distortions and
magnetic ordering effects, we estimate that the $\Gamma_7/\Gamma_8$ CF
splitting is between 73 and 142 meV, in good agreement with the
measured value of 131 meV from neutron diffraction
\cite{Boothroyd2001PRL2082}.

\subsection{Aspherical SIE in other methods}
\begin{figure}[htbp]
\includegraphics[width=0.8 \linewidth]{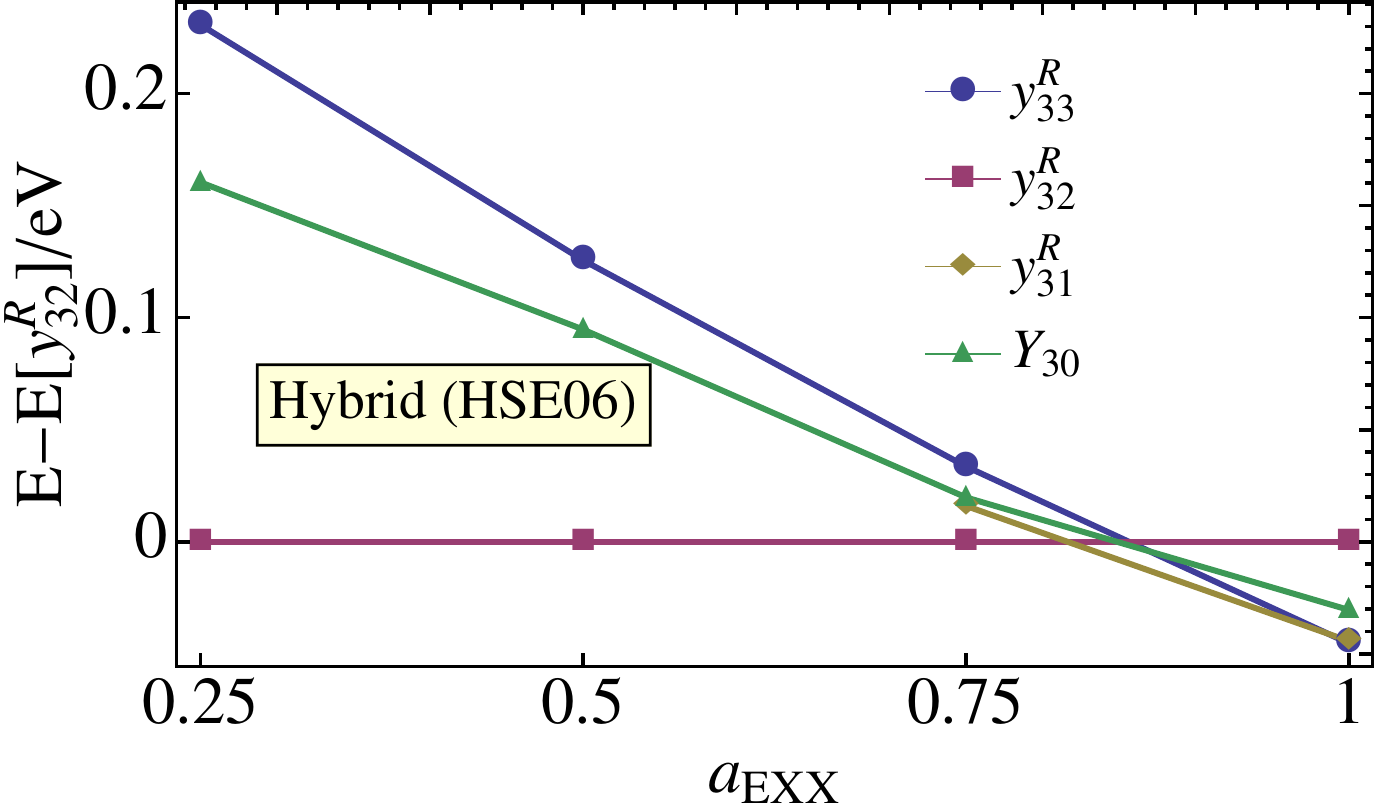}
\caption{Total energy of the Pr$^{4+}$ ion for different filling of real orbitals as a function of $a_{\mathrm{EXX}}$, the fraction of exact exchange, with the hybrid functional (HSE06) method. Most calculations for complex orbitals $Y_{3m}$ converged to very different states and are not shown.}
\label{fig:EvsaEXX}
\end{figure}
%It bears some resemblance to hybrid functional approaches \cite{Novak2006PSSB563}, . 
Our method bears some likeness to the hybrid functional approach. The difference in the latter is that
the exchange interactions are calculated directly from the wavefunctions, with the amount of exact or Fock exchange ($U/2+a J$ for one localized electron in terms of LDA+$U$) as well the replaced LDA/GGA exchange controlled by a fixed parameter $a_{\mathrm{EXX}}$. However, $a_{\mathrm{EXX}}$ in the hybrid functional method is often system-dependent and fitted to experimental data, just like $U$ in LDA+$U$. For instance, Ref.~\onlinecite{Torumba2008PRB155101} found that in $f$-ystems good results were obtained using $40-70$\% Fock exchange, while $d$-systems typically require $20-50$\% \cite{Cora2004SB171}. However, such an $a_{\mathrm{EXX}}$ may not necessarily lead to accurate removal of the aspherical SIE. Fig.~\ref{fig:EvsaEXX} shows the energy of Pr$^{4+}$ ion as a function of $a_{\mathrm{EXX}}$ calculated with the hybrid functional (HSE06) \cite{Heyd2006JCP219906}. Nearest degeneracy is obtained at $a_{\mathrm{EXX}} \approx 85\%$. Given the sensitive orbital dependence of SI demonstrated in this work, in general the accuracy of hybrid functional calculations for $f$-electron systems may still suffer from incomplete removal of aspherical SIE. After the first submission of this manuscript, we became aware that the idea of removing on-site $E_{\mathrm{H}}$ from LDA+$U$ was previously proposed from a different perspective in Ref.~\onlinecite{Seo2007PRB033102}, in which the correction energy is independent of the orbital filling, an important different from our approach. Therefore, the method of Ref.~\onlinecite{Seo2007PRB033102} is not expected to give accurate removal of the orbital-dependent SIE.

\section{Summary}
% FZ: maybe this is a bit long?
% VO How about this? I am not sure I like mentioning the hybrid
% functionals again.
In summary, we have identified a serious problem in applying the
LDA+$U$ method to $f$-electron systems: the degeneracy of atomic
orbitals is lifted, resulting in qualitatively incorrect electronic
ground states and orbital excitation spectra. Aspherical
orbital-dependent self-interaction is identified as the main source of
error. To correct it, a new LDA+$U$ scheme is proposed,
which leaves the Hartree intact and only replaces the LDA
exchange with the Hartree-Fock exchange. Our method has one
adjustable parameter $U$, with the other two ($J$ and $c$) being
determined from the condition of orbital degeneracy in free ions. The
computational expense is approximately the same as in the conventional
LDA+$U$, and very competitive compared to %much less than that required in 
hybrid functional approaches \cite{Novak2006PSSB563}. 
We expect that our
method will scale to large systems and will significantly improve the
accuracy of first-principles studies of $f$- as well as heavy $d$-systems with significant relativistic effects. Additionally, more advanced methods such as GW and DMFT could benefit from the correct input ground state orbitals generated by our method.

% As a solution, we propose to reformulate
%the LDA+$U$ method by leaving the LDA Hartree term intact and only
%replacing LDA exchange with the Hartree-Fock exchange. In our method,
%the Hartree-Fock exchange term cancels the LDA self-interaction energy
%to a high degree of accuracy, ensuring near-degeneracy of real- and
%complex-valued orbitals in free ions and correctly reproducing the
%electronic ground state and excitation energy in the PrO$_2$ solid.
%Even though $d$-electrons do not suffer from the aspherical
%self-interaction error when the orbital momentum is suppressed and the
%wavefunctions can be chosen real, inclusion of spin-orbit coupling
%effects in heavy $d$ elements may also bring in complex wavefunctions
%and associated ordital-dependent self-interaction errors; further
%tests of these effects for $5d$ elements are needed.

\acknowledgements{
This work was supported by the U.S. Department
of Energy, Nuclear Energy  Research Initiative Consortium (NERI-C)
under grant No. DE-FG07-07ID14893. We gratefully acknowledge helpful
discussions with Drs. M.\ Asta,  C.\ Wolverton, G.\ Kresse,
M.\ Cococcioni, M.\ van Schilfgaarde and S.\ Barabash.} 
%\end{acknowledgements}

%\bibliography{../../../Documents/manuscript/newbib,../../../Documents/manuscript/other}

\end{document}